\documentclass[12pt]{article}
\usepackage[pdfstartview=FitH,colorlinks=true,linkcolor=blue,anchorcolor=red,citecolor=magenta,urlcolor=blue]{hyperref}
\usepackage[english]{babel}
\usepackage{amsmath,amssymb,titling,authblk}
\usepackage{slashed}
\usepackage{amsmath}
\usepackage{amscd}
\usepackage[normalem]{ulem}
\usepackage{appendix}
\usepackage{bbold}
\usepackage{bm}
\usepackage{bxcjkjatype}

\usepackage{pdflscape}
\usepackage{yfonts}
\usepackage{amscd}
\usepackage{epsfig}
 \usepackage{microtype}

\usepackage{cite}

\allowdisplaybreaks[4]
\numberwithin{equation}{section}

\usepackage {color}

\numberwithin{equation}{section}
\usepackage{float}
\restylefloat{figure}

\usepackage[T1]{fontenc}
\usepackage{lmodern}

\begin{document}

\setlength{\droptitle}{-6pc}

\title{ Quantum Spacetime: Echoes of basho}

\renewcommand\Affilfont{\itshape}
\setlength{\affilsep}{1.5em}

\author[1,2]{Fedele Lizzi\thanks{fedele.lizzi@na.infn.it, fedele.lizzi@unina.it}}
\affil[1]{Dipartimento di Fisica ``Ettore Pancini'', Universit\`{a} di Napoli {\sl Federico~II}, Napoli, Italy}
\affil[2]{INFN, Sezione di Napoli, Italy}

\date{}

\maketitle

\vspace{-2cm}

\begin{abstract}\noindent
I will discuss how the concept of \textit{basho} (\begin{uCJK}場所\end{uCJK}), introduced by Nishida Kitar\=o nearly a century ago, can give an interesting insight to understand the concept of a point in modern quantum gravity. A quantum spacetime, necessary for the quantization of gravity, requires a whole rethinking of geometry, starting from the primitive concepts, like that of a point. I argue that the local vision of what becomes of classical points in quantum gravity, and in particular in noncommutative geometry, shows several similarities with Nishida's basho. 
 \end{abstract}

\newpage

\section{Quantum Mechanics and Gravity} \label{intro}
Probably the most important unresolved issue in contemporary physics is the coming together of quantum and gravity. Not that we really understood either, in particular the second. Richard Feynman, one of the most prominent theoretical physicists of the second half of the twentieth century,  said ``I think I can safely say that nobody understands quantum mechanics''. Nevertheless we have working theories in both cases. Both quantum mechanics, and its evolution in the quantum theory of fields\footnote{There is a vast literature on quantum mechanics, and quantum field theory, usually technical textbooks. A gentle introduction for philosophers can be found in the site of the \textit{Stanford Encyclopedia of Philosophy}, \url{https://plato.stanford.edu/entries/qm/}, the encylopedia entry has in the end a large set of references.}, and general relativity\footnote{The most authoritative textbook of General relativity is\cite{Misner:1973prb}.} have had spectacular successes in fundamental as well as practical issues. It suffices to mention the Higgs particle, or the precision of the Global Positioning System, better known as GPS, which would be imprecise if the corrections due to general relativity were not taken into account. The two theories, like every theory, have a range of application, i.e.\  they work in a particular interval for the values of energy, or velocity, size, mass\ldots There is a (high energetic) range in which the two theories should both be valid, but cannot be. Stretching quantum field theory in a range in which gravitational effects are important, renders the theory useless from a predictive point of view. The infinities which are present in the theory, but usually are \emph{renormalised} to give a finite answer, are now untamed and the theory is useless. 

While this energy regime is beyond the reach of present (and probably also future) accelerators, it is relevant for supernova explosions, and above all for the primordial universe\footnote{The universe started in a very dense state where matter and radiation (light) blurred their differences, in this ``primordial soup'' both general relativity and quantum mechanics become relevant.}, few instants past the Big Bang. Even more important in this context, is the conceptual issue. Quantum field theory describes with great success the three interactions which govern the microscopic world, while general relativity describes the interaction relevant, for example, for the motion of planets and ping pong balls. If we want to have a full comprehension of the relations among ``things'' we need a theory which encompasses all of the interactions at all scales.

Classical mechanics is the theory of the motion of bodies or particles, described by their position and momenta, which, in the absence of magnetic fields, we may identify with velocity. The set of all possible position and momenta form what is called the \emph{phase space}. In other words the state of a system (particle, extended body, or more complex systems) is described by a point in phase space. In quantum mechanics we cannot anymore describe the system by such a point, the uncertainty principle formulated by Heisenberg, and which is at the roots of quantum mechanics, states that if we measure with great precision the position of a particle, this forces an uncertainty on the measurement of the velocity, and specularly, measuring with precision the velocity renders uncertain the position.
Heisenberg uncertainty principle therefore prevents us from using the geometry of phase space as made of points. Something else has to be invented, a theory of Schr\"odinger probability wave functions, or in the equivalent Heisenberg formulation, a theory of operators (finite or infinite matrices) acting on a Hilbert space\footnote{The original article of Schrodinger is in~\cite{Schrodinger1926}, an English translation of it is in~\cite{WheelerZurek}. The original article of Heisenberg is~\cite{Heisenberg1925}, and its English translation can be found in~\cite{vanDerWaerden}.}. Nevertheless the space of positions is still there. We may perfectly localise a particle if we are willing to renounce to any information about its velocity.

General relativity on the other side is the theory of curved spacetime, it is very much a geometrical theory. Curvature determines the motion of bodies, which in turn deform the curvature. Geometry becomes therefore the dynamical variable of the theory. A quantum theory of gravity will need some form of quantization of spacetime itself. 

Many physicists, mathematicians and philosophers have studied the relations between spacetime and the quantum, and I think that the work of Nishida Kitar\=o can give useful insights in the conceptual framework of a quantum spacetime.

I am neither a philosopher, nor a translator, but I have the impression that, in bringing the work of Nishida to the attention of western researchers (and thus enabling me to read it), one of the problems has been the proper translation of the word \begin{uCJK}場所\end{uCJK}, translitterated as \emph{basho}. In English it is usually rendered as ``topos'' or ``place''. I will follow the second translation, because I like the idea that Nishida's considerations were about a physical aspect, operationally definable, rather than a more abstract philosophico-mathematical concept. Place has immediately to do with space. Something concrete, that we can easily define. More or less. As we will see, much less than more. 

I can say, a little mischievously, that philosophers write books about single (and often illusorily simple) concepts, and for them this it is normal. They know that they do not know, and this enables them to safely write about anything without the need to give precise definition.
Mathematicians write books and papers, about concepts which they decide are worth knowing, whose meaning they define precisely, and in the end all of their results end up being tautological consequences of their definition.
We physicists instead like to think that we write papers about things which we not only know, but which we can measure and even put to good (or sometime evil) use. When probed about the need to actually know what we are doing, we usually ``shut up and calculate''.
In the following I will argue that in physics (and philosophy) we need the concept of place, although we do not know what it means. 

\section{Points and pointlessness \label{pointandpointless}}

When I started to become acquainted with the work of the Kyoto School, and in particular the philosophy of Nishida I was stricken by he parallelism between 
 Nishida's philosophy and quantum spacetime. They both challenge traditional views of space, emphasizing the relational nature of existence, the indeterminacy inherent in reality, and the role of observers in defining the physical world. The similarities (and the differences) suggest that exploring the philosophical aspects of quantum theories could provide valuable insights into the nature of spacetime. In the rest of this section I will describe some of the physical aspects which inspired me in the parallelism. I will start asking a very simple question, what is a point?

Physics is deeply rooted in geometry, which in turn usually studies spaces made of points, and their relations. That of a point is a basic concept, and usually little time is employed to define it. It is  usually encountered in the first lesson of geometry. I still remember that
I first encountered the concept in my elementary school book, where a point was defined as: \emph{A Geometrical Entity without Dimension.}  Over six decades later I am not sure what it means. Euclid  defined a point as \emph{ That which has no part}. 
The highest authority I can think of, \textsl{Wikipedia}, states: \emph{ In Euclidean geometry, a point is a \emph{primitive} notion upon which the geometry is built.} 

The key word in the previous definition is \emph{primitive}. It exonerates us having to define it. But in physics we like to operationally define it in some way. According to the theory we are considering, a point may be the possible position of a \emph{point particle} in space, or the combination of its position and velocity in phase space. The position (and possibly velocity) of an ensemble of particles is a point in a space with many dimensions\footnote{Following the use among physicists I use the word dimension with a variety of different meaning. There is the dimension of a space, in which sense a solid ball is three dimensional, a plane is two dimensional, a line is one dimensional and a point is zero dimensional. Then I use it to indicate the nature of the physical quantity (occasionally this is called engineering dimension), for which a rod has the dimension of a length, and I can measure it in meters or any other unit of measure, a time in seconds, a velocity in meters per second or Kilometres per hours. I avoid the third possible use, to indicate how large is something, using instead ``size''.}. It may happen that the number of dimensions becomes too large, for example if we are describing a gas with a very large number of particles, each counting for six dimensions. We are then forced to drastically reduce this number, and consider a much reduced number of degrees of freedom, making averages. The state of a gas is efficiently described using pression, volume and temperature, so that it becomes a point in a three dimensional space. 

Let us  pause for a moment and consider this. The knowledge of the temperature, pressure and volume of a gas does not specify the state of its molecular components. Many different configurations would give the same state in this reduced space. Indeed one of the basic principles of thermodynamics is that this is a state which can be achieved in the largest possible number of equivalent configurations of microstates. Nevertheless in classical mechanics there is, for a given gas at a given time, \emph{the} configuration of the single components. It is technologically impossible to measure this set of numbers. There would be something of the order of $10^{23}$ numbers, and it is unthinkable just to think of storing them. Writing one hundred of them on a single sheet of paper, and piling them up we would get a tower of about a billion light years. If you are more technological, then one thousand billions of terabytes will suffice. An average hard disk of one terabyte is about a centimetre tick, then piling up the lot, a mere tenth of a light year will be enough... All this for a single instantaneous configuration! Nevertheless, in classical mechanics, the configuration is there, and the fact that we do not measure it does not signal a conceptual limitation. 

But if we limit ourselves to a small number of particles, one for example, its position in space, or phase space, is classically well defined, and we may think of measuring it with arbitrary precision. A point particle will define a point where it sits.

Before Einstein, localisation in space (points) was conceptually very different from that in time (instants). Relativity fused space and time into spacetime, and the notions of point (in space) and instant merged into the \emph{event}. This is a revolutionary change of paradigm\footnote{It is worth noticing that the concept of spacetime implies a dual complementarity of the points and instants, and this was inspiration for Nishida's 
leap of linking consciousness (a temporal characteristic) to the spatiality of the body. I thank Prof.~Tremblay for bringing this aspect to my attention.}, but relativistic geometry is still the usual one, based on points which do not radically differ from the ones analysed previously. For this reason most of the considerations which follow are valid both in space or spacetime.

So far I talked of place and point in an interchangeable way. Confining ourselves to the geographical frame and to language, a place (another primitive entity) is more than a point. Not only it may be an aggregate of points. We say:  \textit{It is a nice place to visit}. It has an implicit relational meaning. In some sense for a point to be a place something should happen there. The relational aspect is also central in basho.
 Promoting the points/instants to events is not only (an already momentous) enlargement of space to spacetime, but implicit in the concept of event there is the possibility that something happens \emph{here and now}, or, since spacetime is the set of all events, there and then. Not only this, but events are always relative to an observer, a reference frame\footnote{Although the two concepts of observer and reference frame have subtle differences, I will use the two terms interchangeably here.}, and this implies the existence of an entity external with respect to the event in question. Since quantities that were considered to be absolute (space, time, speed, \ldots) have become relative, we should always state (often we do it implicitly) to which reference frame/observer they relate.

We are assuming that it should be possible to measure, in some way, the event. And this means that operationally we have some meaning to measure, to ``to know'', about it. In their evolution from Newtonian to relativistic physics, points move from subjects to predicate. They are not the assumed primitives, and therefore not defined, elements of the space, but have become objects which emerge from relations, are possible container of information. In order to talk of points we need these extra structures. In the physics literature there is no distinction between point and place, the latter in fact is seldom used. We call point the same entity, from Newtonian physics to quantum gravity. But the concept evolved dramatically. In Nishida the two concept are more clearly distinguished in the duality subject/predicate.
The fact that points (or places) are predicates is one of the tenements of basho. 
Let me quote Nishida: \textsl{``that which becomes subject and does not become predicate – that which becomes predicate and does not become subject.''} (letter from Nishida to Tetsugaku Kenky\=u, my translation from~\cite[Pag.~4]{Fongaro}). Here, Nishida discusses the dynamic interplay between the subject and the predicate, emphasizing the philosophical significance of their relationship in the context of his broader metaphysical framework. This led me to the statement made in the introduction that rather than points, in modern physics, we should talk of basho. In the following I will further analyse the modern concept of point, arguing that in the physical quantity we call with the same name, the duality is already present.

By the basho, Nishida explores the significance of place in understanding reality, emphasizing the importance of relational qualities. On the other side, I hope I made it clear that quantum spacetime theories challenge traditional notions of space by replacing points with noncommutative algebras or other structures. The idea of a point in space loses its meaning below certain scales, aligning with Nishida's emphasis on the relational aspect of existence. This is a non-substantiality and relational nature of things, and in particular of geometry.

Theories in physics are usually required to be local. This means that we have to respect the localisation of our entities, their position in space. Classically we are dealing with points in phase space, let us say determinations of position and velocity. \emph{Quantum Mechanics} arrived  more or less at the same time as special relativity (a few years later in its modern formulation), and it was a complete game changer.  And the first example of a Noncommutative Geometry, which we will discuss later with its application to spacetime.

There is an important aspect of quantization of spacetime which I want to stress: the impossibility to define a point in phase space due to Heisenberg principle. Our space is now the phase space of a single particle in one dimension\footnote{At the time of Nishida the usual interpretation was that of Bohr complementarity, for which an electron sometimes act as a wave, sometimes as a particle. It became clearer and clearer that the electrons behave always as electrons. It is us that, based on the similarity of the equations,  sometime find it useful to compare them to the wave of water, and sometimes to tennis balls.}.
And to further simplify I often identify momentum with velocity, which is correct if there are no magnetic fields, otherwise things are more subtle. These simplifications do not alter the conclusions. A subspace of phase space is configuration space: the space of all positions. Classically phase space is a geometrical entity, technically the cotangent space of the configuration space. Classical mechanics is in some sense the study of this space, and how its points evolve. It usually has further geometric characterisations and structures studied in differential geometry.  In this sense the space may be curved \footnote{This means that for example in two dimensions a plane sheet of paper tangent to a point will change its orientation, and this generalizes to higher dimensions.}, distinguishes between position and velocity\footnote{This is captured by a differential form called \emph{symplectic form}.} in a way which does not depend on the coordinates, or how the phase space changes when the particle is subject to constraints, for example is forced to move on the surface of a sphere.

All this collapses because quantum mechanics denies the very existence of the points over which the geometry of phase space is built.

\begin{figure}[htb] \centerline{\includegraphics[scale=0.5]{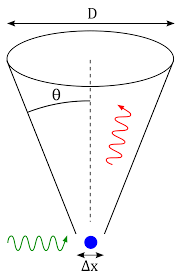}} \caption{ A schematic view of the Heisenberg microscope. Picture taken from wikipedia \url{https://en.wikipedia.org/wiki/Uncertainty_principle} under Creative Common License \url{ https://creativecommons.org/licenses/by-sa/3.0/ }}  \label{Fig1} \end{figure}

Let me give a very heuristic illustration of the Heisenberg's principle, the so called \emph{Heisenberg Microscope.} Point are the set of all places in which a particle\footnote{Indeed we usually talk of \emph{point} particle.} can be. In order to see that a particle is somewhere, to measure its position, we should shine some light, or other form of radiation on it. If the particle is there the particle will scatter it, and we can see it with a device, some sort of microscope. This is an ideal experiment, and the process is illustrated in Fig.~\ref{Fig1}.

The problem is that \emph{point radiation} does not exist. Any radiation is a wave and it has a finite wavelength, $\lambda$, and therefore the measurement will have an uncertainty $\Delta x$ given by this wavelength. One can however consider radiation with shorter and shorter wavelength, to improve the measurement so that the uncertainty becomes smaller and smaller ($\Delta x\to 0$) so that the measurement becomes more and more precise.  Here enter the new actors, relativity and quantum mechanics, and their introduction of two fundamental constant, the speed of light $c$ and the quantum of action $\hbar$. According to relativity (it is a celebrated result by Einstein himself) light comes in quanta of radiation: \emph{photons}. Any photon has an energy proportional to the inverse of its wavelength: $E=\frac\hbar{2\pi\lambda}$. Therefore a ``small'' (short wavelength) photon, necessary to measure the position with high precision, will be very energetic. In the interaction with the particle some part of this energy will be transferred to it, but since any microscope, even ideal, has a lens with finite diameter, there will be an uncertainty in the fraction of energy transferred, in the form of velocity of the particle after the collision. Hence knowing with precision the position implies an uncertainty in the velocity (momentum). This is the content of Heisenberg's uncertainty principle. 

The reasoning is totally heuristic, and in fact, doing properly the calculation, the limit one finds to the product of uncertainties is off by a factor of $4\pi$. The proper way to see the principle uses the full formalism of quantum mechanics. Position and momentum, become different mathematical objects, noncommuting operators on a Hilbert space\footnote{A Hilbert space is an infinte dimensional generalization of the usual space of vectors, like three dimensionals vectors in space, i.e.\ pointed arrows in space.}. The state of a particle is described by a wave function, measurements are probabilistic, and we need a whole new paradigm. The physics of quantum phase space for nonrelativistic point particles is relatively well known. The founding fathers: Bohr, Schr\"odinger, Heisenberg, Dirac, Pauli~\ldots based on the mathematics of  Weyl, Hilbert, von Neumann and others, gave us what we call quantum mechanics. I dare not affirm that quantum mechanics is well understood, but I venture to say that we have a solid mathematical framework with which we can work, often with spectacularly accurate predictions.

Another important aspects of quantum mechanics concerns us:  the presence of \emph{observers} as fundamental actors in the play. A physical state looses his absolute meaning, it is such in relation to observers. The famous example of the Schr\"odinger cat shows this in a dramatic way (especially for the feline). But also our previous discussion of the Heisenberg microscope shows that, quantistically, geometry has its physical meaning when probed, when observed. The role of the observer is central in nearly all of the interpretations of quantum mechanics, including of course Nishida's. The relational aspect between point and observer becomes fundamental. Generalising the location in a point to the more general concept of state, this is one of the central principles behind Rovelli's relational vision of quantum mechanics\footnote{Relational Quantum Mechanics suggests that the properties of quantum systems are not absolute, but are meaningful only in relation to other systems. Quantum states and events depend on interactions between the observer and the observed, implying that different observers can have different, equally valid accounts of the same event. This interpretation aims to address foundational issues in quantum theory by eliminating the need for an observer-independent state.~\cite{Rovelli}.}.

\section{From Points to basho}

Another central aspect in quantum mechanics, which is playing an increasingly important role in quantum gravity, is the role of the observer, and of the reference frames. Much of the confusion about quantum aspects stems from the difficulty of properly defining and understanding the role of the observers. I personally think that Feynman was thinking about these aspects when he said the sentence I wrote at the beginning. In quantum gravity, it is now becoming clear that one has to quantize observers/reference frames as well~\cite{LMMP, Giacomini}. This implies the possibility of superimposed and entangled observers. Moreover, in quantum gravity as in quantum mechanics, the very act of measurement affects the outcome, and the concept of observables becomes central in the act of measurement, not as a passive spectator, but an active actor. The relational nature of spacetime is closely tied to the perspective of an observer. This is very much present in Nishida's work; he introduces the idea that spatial things must be more than spatial, and their existence is recognized through qualities that participate in the place where they are. This suggests an observer-dependent aspect of reality very much similar to what is going on in quantum gravity, several decades after he wrote his papers!  In this sense Nishida can be seen as a precursor of what I call ``the information frontier'' of quantum gravity, namely the recent realizationn that quantum information should play a fundamental role also in quantum gravity, beyond the foundational and computational aspects.

In quantum gravity, the act of measurement is not just a passive observation but an active interaction that shapes the reality being measured. This resonates with Nishida's idea that the act of knowing (epistemology) and the nature of being (ontology) are intertwined. In the quantum realm, observers are part of the system they measure, influencing outcomes in a way that classical observers do not. This observer-dependent reality is reflected in the relational properties of noncommutative geometry, where measurements are defined not by fixed points but by the relationships between different algebraic elements.

Nishida's basho is very much in this relational line, even if he seems to point to something behind it: \emph{``But if we are to think in
such terms, the related and the relation must be one. It would be, for
example, like physical space. However, that which relates physical space to
physical space is no longer physical space, and there must further be a basho
wherein the physical space is implaced''} \cite[Pag.~50]{placedialectic}.
Nishida was talking of  the physical space of his time, not exactly the one I will describe later. Nevertheless even in this case his intuition was that the spaces of physics cannot be considered without relations to other agents. In this case I do not interpret the ``further basho'' in which the physical space is implaced as a superior geometry, but rather as a more general framework in which the theory must live. I will given a pointer to the ``further place'' in the following. Nishida's idea parallels the view that physical space is a manifestation of more fundamental, relational structures in quantum theories, providing a philosophical underpinning to the abstraction in modern physics.
At any rate, despite the impossibility to define points in the whole phase space, in the configuration space (which is a subspace of the former), points are a well defined concept in nonrelativistic quantum mechanics.

It was clear from the beginning that quantum mechanics should encompass relativity. For the time being just consider special relativity; general relativity and gravity will be treated  later. Relativistic quantum mechanics is not a well defined theory. The problems stem exactly from the geometry of spacetime: the very concept of position is problematic. The association of well defined operators to the usual coordinates of spacetime is not well defined. It is necessary to stretch things a great deal.  There are some sort of coordinates (Newton-Wigner) which can be defined but they are exceedingly contrived. Relativistic quantum mechanics, as described by the Dirac equation, although it had important early successes (like the hyperfine structure of the hydrogen atom), shows several inconsistencies if merely seen as a relativistic version of the Schr\"odinger equation.

The necessary further step is to proceed from quantum mechanics to the quantum theory of fields. The theory is now relativistic (in the sense of special relativity), fields are defined on spacetime, and 
the local aspect of the fields plays a fundamental role~\cite{Haag, Weinberg}. Quantum relativistic fields are functions which, in some sense,  generalise the idea behind Schr\"odinger wave functions. A relativistic theory will have Lorentz invariance, i.e.\ it will described in the same way by observers rotated or in relative non-accelerated motion with respect to each other.  
Together with Lorentz invariance, which describes transformations, there is \emph{locality}. Points have the potentiality to be events, and it is fundamental to give them a causal ordering. The geometry of spacetime is fully integrated in the theory, but the geometry of spacetime is the one of non quantistic (but relativistic) theories. Fields depend on the spacetime coordinates, as well as other possible parameters (the spin of the particle for examples), but these position are not operators, as in the nonrelativistic case. Time, which never was an operator, is on a par with the other spatial dimensions, as it befits to a relativistic theory. Of course it is not that we have gone back to classical mechanics: the objects which are now operators are the fields themselves. The quantum theory of these fields has had spectacular successes for the understanding of fundamental interactions, the already cited Higgs being only one of these.

Nishida argues that what ``is in'' something must have a relational connection to the place where it exists. The essence of a thing is recognized through its relationship with others. This is a general feature, but it becomes poignant when related to locality. Quantum spacetime theories also challenge the idea of substantial, well-defined geometrical entities. The very concept of points or locations becomes blurred, and the focus shifts to relational/global aspects. The shift goes to functions, fields, which are extensive quantities that involve geometry globally.

Note that so far we had used only two of the fundamental constants of nature: the speed of light $c$, and Planck's constant $\hbar$. In the heuristic Heisenberg microscope the speed of light appears in a rather indirect way, it is the parameter necessary to write Maxwell's equations, which describe the behaviour of radiation. Instead $\hbar$ appears more directly in the relation $\frac\hbar{2\pi\lambda}$. Let me comment a little on these constants. 

A quantity with the dimension of a velocity, i.e.\ a length divided by a time, is necessary for the very definition of spacetime. This is the space identified by four coordinates. Three of them are the usual $x,y$ and $z$, or $x_1, x_2, x_3$, but in order to have the fourth, with the dimensions of a length, related (proportional) to time, we need a quantity which, multiplied by a time, gives a length: a velocity. The fundamental intuition of Einstein was that this quantity should be independent on the reference frame, and this is one of the main pillars of relativity. With $c$ we may define the fourth coordinate $x_0=ct$, with the dimension of a length, on the same footing as the other three. 

A similar role is played by Planck's constant with respect to phase space. It is usually said that the dimensions of $\hbar$ are that of an energy times a time, or a mass times a length squared divided by a time, but these dimensions can be equally seen as that of a length times a momentum (in turn a mass times a velocity). Therefore $p\sim mv$ has the same dimensions of $\hbar x$ (in one dimension). This gives full geometrical validity to phase space. We see that in both cases, already at the initial level of definition, the possibility of geometrize spaces gives points which should always be defined in a relational sense.  

Quantum field theory describes only three of the fundamental interactions, it does not describe gravity. The theory which describes gravitation exists, it is called \emph{general relativity}, and it is the theory of a \emph{curved} spacetime. While it is possible to have a quantum field theory in a curved spacetime, seen as a background, this is not a quantization of gravity. In general relativity the curvature of spacetime is \emph{dynamical}, it is not anymore the arena in which the physical quantities live, but the dynamical variable of the theory. The presence of masses alters spacetime (its curvature), and the curvature in turn determines the behaviour of the masses. The difference is subtle but momentous. Rather than having the earth going around the sun in spacetime, there is a system of earth and sun which curve space, which forces the two celestial bodies to follow a trajectory, and while they move they deform, curve the spacetime. The equations are such that the Kepler orbits are reproduced, with some small differences, which can be measured with high precision, for example in the case of the precession of the orbit of the planet Mercury.

Spacetime has become therefore \emph{the} dynamical variable of the theory. There is not anymore just a system of particles which moves, evolves, in spacetime. The object which evolves is spacetime itself. This forces again a rethinking of geometry. If we attempt to define points in space(time) at very short distance we run into trouble if we put together quantum mechanics and gravity. This is similar to what we discussed for phase space, but this time the subject of the discussion is spacetime itself, the concepts of ``here and now''. I will use a reasoning very similar to the Heisenberg microscope described earlier. It was proposed for the first time by Bronstein~\cite{Bronstein} in 1938, but presented independently in a modern and most terse way, by Doplicher, Fredenhagen and Roberts~\cite{DFR} in 1994. I will present a caricature of these arguments, which however captures the main idea in a nontechnical way.

We are now interested only in space, and not momentum. In quantum mechanics there is no limitation to an arbitrarily precise measurement of $x$, if we are not interested in the velocity aspect. We can therefore use a small probe, as small as we want. It will be very energetic, but this is of no concern to us now, we just want to know if there is an electron at a given point. We have to concentrate an arbitrarily high quantity of energy in a small region. This is not a problem in quantum mechanics if, as I said, we do not care of the velocity afterward. But in general relativity energy is the same as mass, and concentrating an high mass in a small volume curves spacetime. 

Ultimately, if too much mass is concentrated in too small a volume a \emph{black hole} will develop. This is not one of the cosmic objects whose gravitational waves are now being detected, with a mass several times the mass of a star. It is rather a micro-black hole, which nonetheless develops an horizon, and no information can exit this horizon. The scattered photon indicating the presence of a particle at a point would not reach us, and therefore probing sizes below a certain scale is impossible. 
It is possibly (ideally) to detect the black hole, but not to ``see'' anything inside its horizon. Again there is a limit to the precision of the measurement.

This is occurs because gravity introduces another fundamental constant, Newton's constant, the quantity which appears in the universal law of gravitation which states that two massive objects will attract each other with a force of magnitude proportional to the product of the masses $M_1$ and $M_2$, and inversely proportional to the square of the distance $r$:
$$
F=G\frac{M_1M_2}{r^2}
$$
Newtonian gravity lives in an absolute space, and in it $G$ plays a dual role, on the one side it is a \emph{coupling constant}, i.e.\ it gives the strength of the interaction between the bodies, but on the other hand it gives a scale. Combining it with the other two constants, $c$ and $\hbar$,  we can define a quantity with the dimension of a length: $\ell=\sqrt{\frac{\hbar G}{c^3}}\simeq 10^{-33} \mathrm{cm}$, about a billionth of a billionth of a billionth of a millionth centimetre. This quantity is called \emph{Planck lenght} and gives a fundamental scale to spacetime.  Doing the calculations for the black hole discussed above, its size results indeed to be this Planck's length.

In the discussion of the ``Bronstein microscope'' I have been very heuristic, as in the Heisenberg case. To make rigorous statements, it would be necessary to have a full theory of quantum gravity, a theory which we do not have. I (perhaps optimistically) like to think that we do not \emph{yet} have. The search for quantum gravity is very actively going on in the physics community. There are various approaches to it, not necessarily completely alternative to each other. I may mention strings~\cite{Polchinski:1998rq,Polchinski:1998rr}, loop quantum gravity~\cite{RovelliLQG}, asymptotic safety~\cite{Percacci}, and noncommutative geometry~\cite{WSS}.

As we said earlier, spacetime in general relativity is the dynamical variable of the theory. Therefore, quantizing gravity must mean quantization of spacetime; we are in the presence of a quantum geometry of spacetime. This has analogies with the quantization of phase space, but the object we are required to quantize is now the set of all possible positions, of all places in which things can be.

The two ideal microscopes, Heisenberg and Bronstein, that we presented, and more profoundly the formalism of quantization, have an important and central aspect in the uncertainty or indeterminacy of central local objects. Here I may mention another aspect, which I have not treated here, which is peculiar to quantum mechanics: \emph{superposition and entanglement}. Quantum states can be in a superposition which renders them indistinguishable: spin up or spin down, two positions, to the extremization of cat dead/alive. In Nishida, this is reflected in his acknowledgment of the indeterminacy of existence and the challenge of defining things precisely. But since we can only have simple singular measures, there is an implicit implication of a level of indeterminacy. Similarly, quantum spacetime theories, such as noncommutative geometry, show that precise localization below a certain scale becomes operationally undefinable. The uncertainty principle becomes a fundamental pillar of the construction in both contexts.

I have treated extensively the philosophical implications of noncommutative geometry in a paper with Huggett and Menon in \textsl{Synthese} titled ``Missing the Point in Noncommutative Geometry''~\cite{HuggettLizziMenon}, and I refer to this paper for more details. The central point of noncommutative geometry is a change of perspective. Usually, one starts from the points, and then further structures are defined on them, like functions, i.e.\ the assignment of a numerical value to each point, for example, the value of the $x$ coordinate, or the intensity of the electric field. Noncommutative geometry dualizes this. To start with, we have functions, and how these combine among themselves (technically they form an \emph{algebra}). If the functions commute among themselves, that is, if the order in which they are multiplied is the same ($f$ times $g$ is the same as $g$ times $f$), then it is possible to reconstruct a usual space of points and to describe all geometrical quantities using relations among functions. If the functions instead do not commute, then it is no longer possible to have points, but several higher structures survive, and we have a noncommutative geometry. This is what happened for the geometry of the phase space upon quantization; this is what could (and I believe should) happen for the quantization of spacetime. Clearly, at some stage, in some limits (long distances or low energies), something that resembles points should exist; otherwise, we would be forced to deny all the successes of physical theories, with their predictive power. These theories then become but a particular approximation of a higher theory.

\section{Points, Basho and Measurement}

The points of noncommutative geometry therefore end up being more similar to Nishida's basho than to the points of classical theories. I will take a \emph{tempered operationalist} stand: for me, knowable physical concepts are those for which a measurement procedure is possible. The tempered mitigation is in the possibility that the procedure may be ideal. In this sense, there are no points in spacetime, but only approximations to it, for example, coherent states, which can be seen as probability densities that approximate the concept of a point, but which also require an observer structure. In~\cite{HuggettLizziMenon} we referred to such concepts as operationally definable. To give operationalism substance, one has to specify what measuring operations are available; since we are interested in the possibility of operationalizing points of space.

In physics measurements always come with an error, even if we eliminate all sources of mistakes or systematic errors (like for example a non perfectly calibrated device), there still will a probabilistic error. There is not such a thing as an absolute quantity. This is because any measuring device is affected by implicit, technological limitations, and therefore every measure has an error, which changes from measurement to measurement. The difference between classical and quantum is the fact that in quantum spaces, the uncertainty is inherent in the theory, it cannot be reduced by, say, an improvement of the measuring device. The error cannot go below a certain level. The simultaneous measurement of position and velocity is typical, but there are other examples, whenever the operators connected to the quantity to measure do not commute among themselves.

The limits of measurement are an unavoidable feature of the theory itself,  they are consequences of presence of scales. A measurement implies the presence of an observer and a reference frame (coordinates, units, etc.). In~\cite{HuggettLizziMenon}, we described in technical terms some idealized measurements, showing how, in a particular noncommutative geometry that appeared in string theory~\cite{SeibergWitten}, points, although necessary at the beginning to define the theory, become an untenable concept. We showed how noncommutative geometry gives a complete change of paradigm, from a geometry based on points to a geometry based on algebra, functions, and relations.

Once we have translated geometry into its algebraic counterpart, quantization consists of considering a noncommutative algebra and repeating the analysis, \textit{mutatis mutandis}. In the end, we can show that points are not operationally definable because of Heisenberg's uncertainty! But the rest of the framework remains, and the usual tools used to make physics are still there, except that they must be considered in terms of the generalized functions of the noncommutative algebra. Consider, for example, the central tool of dynamical systems, the action. This is the integral of a function, called the Lagrangian, which can be used to define the equations of motion, which in turn define the time evolution. In classical or quantum mechanics, from the Lagrangian, a new function can be defined, the Hamiltonian, whose quantum version is used in the Schrödinger equation, together with the derivative acting on the wave functions. In the absence of points, we cannot follow the same scheme: integrals are sums over the values of the functions on the points, and derivatives tell how the values of the function vary under an infinitesimal change of the points. This scheme is impossible to implement in a noncommutative geometry, but it can be translated in terms of properties of operators (infinite matrices), so that integrals and actions become sums over diagonal elements of the matrix (this is called the trace), and derivatives become products and differences between matrices. This is a generalization of the matrix mechanics of Heisenberg. It may be worth commenting that there have been applications of this scheme also in the standard model of fundamental interaction among particles and Higgs physics~\cite{allthat}.

It appears that we ended up in a circular reasoning. I started with a space made of points and then showed that the points do not exist in this space. Yet without the initial use of these points, it would be impossible to even start. I stress again that points are not a symbolic device made to explain things, to reduce things to a known situation, such as using ocean water waves to explain electromagnetic (or even probability) waves. Points belong to the ontology of the system, in the sense that they are entity necessary for the setting of the physical model, yet they do not exist; they exist only as a shadow, an approximation.

Another crucial ingredient is the measuring device, which is a generalization of the point itself. A measurement is a way of containig a point into a measurement device, following the analogy I have been pursuing here it is like a basho containing another basho, although of course I am probably stretching the concept well beyond Nishida's intentions. In my operationalist point of view, points, or whatever approximate them, are always a relation between a space and a measuring device. Yet, while we need points, we know that the novel paradigm necessitates their negation in favor of other concepts. To quote from~\cite{HuggettLizziMenon}:
\textit{In theories of noncommutative space, we again assume - but this time for reductio - that there is an ontic state corresponding to an arbitrarily precisely localized particle. We construct the analogue of an epistemic state: a density operator. We then attempt to localize this epistemic state to an arbitrarily small area and discover that this leads to ascriptions of negative probabilities. Since these measures are not elements of the state space, this signals a pathology. The only way to avoid this pathology, we argue, is to drop the assumption that there is an ontic state corresponding to an arbitrarily precisely located particle. Thus, even in principle, it is not possible to localize a particle below a certain area. Operationally, then, such areas - and a fortiori points - are undefinable.}

Furthermore, this approach ties deeply into the relational nature of physical reality as posited by Nishida. In his philosophical framework, Nishida emphasised that entities are fundamentally relational, existing only in relation to a greater context. This perspective aligns remarkably well with the principles of noncommutative geometry, where the traditional notion of isolated points in space is replaced by relational structures. The algebraic structures in noncommutative geometry serve as a mathematical embodiment of Nishida's philosophical idea that space itself is relational.

After having described in a (probably exceedingly short) way noncommutative geometry and quantum spacetime, let me now present a series of parallelisms between these physical concepts and Nishida's philosophy. I have no pretension of completeness, even less of rigour, but reading Nishida, I could catch some aspects of his philosophy which seem to point in the direction I have been pursuing.

From a point of view more centred on philosophy, ontology, and epistemology, some final comments are in order. Nishida's philosophy blurs the distinction between ontology and epistemology. The recognition of existence is intertwined with relational qualities and the act of knowing. There is a perfect parallelism that I can see with quantum spacetime theories, and especially those involving noncommutative geometry. Also, in this case, we have a blurring of the lines between ontology and epistemology. Not only do we somehow need points as a starting point to describe a theory that denies them as fundamental, but also the operational definition of points and the uncertainty principle connect the nature of existence with the act of observation.

\subsection*{Acknowledgments} 
I acknowledge support from the INFN Iniziativa Specifica GeoSymQFT, 
and from Grants No. PID2019–105614 GB-C21 and No. 2017-SGR-929, support by ICSC - Centro Nazionale di Ricerca in High Performance Computing, Big Data
and Quantum Computing, funded by European Union - NextGenerationEU. I would also like to thank Rossella Lupacchini for introducing me to the philosophy of Nishida Kitar\=o and the Kyoto school, and for inviting me to give a talk at the conference \textsl{Philosophy and Physics between Europe and Japan}. This contribution is an elaboration of the material of that talk. Enrico Maresca, Rossella Lupacchini and Jacynthe Tremblay for a critical reading of the first draft of this contribution, their comments and questions, made from the point of view of a philosopher,  helped shape this contribution. I thank Enrico Fongaro for an excellent Italian translations of the work of Nishida~\cite{Fongaro}, which has been the source of my learning about basho. Finally I thank Rossella Lupacchini and Jacynthe Tremblay for giving me the opportunity to contribute to this volume.

\end{document}